# Sub-10 nm precision engineering of solid-state defects via nanoscale aperture array mask

*Tae-yeon Hwang,[1,*] Junghyun Lee,[1,*] Seong-Woo Jeon,[1] Yong-Su Kim,[1,2] Young-Wook Cho,[3] Hyang-Tag Lim,[1,2] Sung Moon,[1,2] Sang-Wook Han,[1,2] Yong-Ho Choa,[4] and Hojoong Jung [1,†]*

[1] Center for Quantum Information, Korea Institute of Science and Technology, 5 Hwarang-ro 14-gil, Seongbuk-gu, Seoul 02792, Republic of Korea

[2] Division of Nano & Information Technology, KIST School, Korea University of Science and Technology, Seoul 02792, Republic of Korea

[3] Department of Physics, Yonsei University, 50 Yonsei-ro, Seodaemun-gu, Seoul 03722, Republic of Korea

[4] Department of Materials Science and Chemical Engineering, Hanyang University, Ansan, Gyeonggi-do 15588, Republic of Korea

* These authors contributed equally to this work.

† Corresponding Author : E-mail: hojoong.jung@kist.re.kr

**Abstract**

**Engineering a strongly interacting uniform qubit cluster would be a major step towards realizing a scalable quantum system for quantum sensing, and a node-based qubit register. For a solid-state system that uses a defect as a qubit, various methods to precisely position defects have been developed, yet the large-scale fabrication of qubits within the strong coupling regime at room temperature continues to be a challenge. In this work, we generate nitrogen vacancy (NV) color centers in diamond with sub-10 nm scale precision by using a combination of nanoscale aperture arrays (NAAs) with a high aspect ratio of 10 and a secondary E-beam hole pattern used as an ion-blocking mask. We perform optical and spin measurements on a small cluster of NV spins and statistically investigate the effect of the NAAs during an ion-implantation process. We discuss how this technique is effective for constructing a scalable system.**

A negatively charged nitrogen vacancy (NV) center in diamond, which is a point defect where two adjacent carbon atoms are substituted with a nitrogen atom and a vacancy, has been recognized as a promising platform for building a solid-state-based qubit system, and it is believed to have the potential to facilitate a variety of applications from room temperature quantum sensors[1, 2] to quantum information processing[3-5]. One of the key advantages of the diamond system is its utilizable interconnection between different atomic spin defect species[6-9], from electronic spins to isotopic nuclear spins. Owing to their high magnetic moment, electronic spins interact with each other with considerably higher dipolar coupling strength compared with nuclear spins, and therefore, they are more suitable for constructing a scalable qubit system[10, 11]. Furthermore, unlike nuclear spins, the formation of electronic spin defects can be tailored with precision, and cutting-edge techniques have been developed to achieve deterministic positioning[12, 13].

The strong coupling between electronic spins has been used to actively study many applications across a range of fields from quantum sensing to quantum information. For sensing applications, dense uniform arrays of ensemble NV spin sensors can improve the magnetic field's spatial resolution in wide-field imaging of magnetic fields[2]. Electronic spins can connect nuclear spin cluster nodes consist of $^{14}$N and $^{13}$C nuclear spins[6, 7], and this fact can be exploited to construct a scalable quantum processor[14]. A strongly interacting electronic spin ensemble can also serve as a good platform for the study of quantum simulations of many body physics[15, 16].

For the aforementioned applications, apart from the fabrication of qubits with good spin properties, the precise controllability of spatial positions, close proximity between qubits for strong coupling, and scalability with uniform coupling strengths are key factors to consider for solid-state based qubit engineering. In particular, the realization of spin-spin magnetic dipolar coupling strength on timescales shorter than the spin coherence time is essential. Therefore, as the size of spin qubits in a system increases, both strong coupling and a long spin coherence time are required. In NV diamond system, for the formation of any non-classical correlated states between NV spins, a 10–30 nm separation distance is typically required[17]; the separation distance is limited by the spin coherence time (Hahn-echo coherence time $T_{2,Hahn}$) at room temperature. In other words, the location precision of NV spin should be less than the sub-10 nm length scale for the construction of a scalable qubit system, which remains as a significant challenge.

For the achievement of such location precision, various methods have been reported to control the position and distance between NV centers in diamond, and a low-energy nitrogen ion implantation process has become an essential tool for the deterministic generation of NVs[13, 18-25]. These methods can be classified into molecular ion implantation[21, 22], single ion implantation via atomic force microscope (AFM) tip hole[18-20], and ion implantation with a blocking mask[13, 23-25]. In molecular implantation, an ionized species such as $N_2^+$ or a molecule with multiple nitrogen atoms[21] are implanted into diamond. The separated nitrogen atoms can be located over a distance range of few nanometers. This method has the advantage of generating closely located NVs with high probability, yet the extent to which scalability can be increased is limited by the number of nitrogen atoms in one molecule. On the other hand, single nitrogen ions ($N^+$) can be implanted into a diamond's surface by using either an AFM tip hole or open apertures with a blocking material covering the area that is not of interest on the diamond. In terms of scalability, the nanofabrication-based mask implantation method is advantageous. While recent improvements have reduced the size and gap between the mask apertures, yet the 2D translational confinement area of the aperture is still too large for realizing uniform strong spin couplings.

In this letter, we present an array of cylindrical apertures with sub-10 nm size and spacing with a high aspect ratio, and successfully demonstrate that nanoscale aperture arrays (NAAs) can be used as a blocking mask during ion implantation. Adopting a cutting-edge nanowire fabrication technique[26, 27], we grow phase-separated eutectic Al-Si thin film directly on a diamond surface. After selective aluminum (Al) etching, we form arrays of nano-porous structures with an aspect ratio of 10. By optimizing the fabrication parameters, we minimize the aperture size to 5.87 nm and the aperture-to-aperture distance to 10.7 nm. To evaluate the feasibility of this approach, we apply a secondary patterned mask on the NAAs through electron beam lithography (EBL) to form small clusters of NVs. This double-layered mask helps reduce the size of the apertures and the spacing between them with a high aspect ratio, more than what state-of-the-art EBL[24] can achieve. Finally, we implant nitrogen ions through a homemade nano-mask and verify the generation of single to triple NV centers by using a second-order correlation function ($g^{(2)}$) and optically detected magnetic resonance (ODMR) signal measurements.

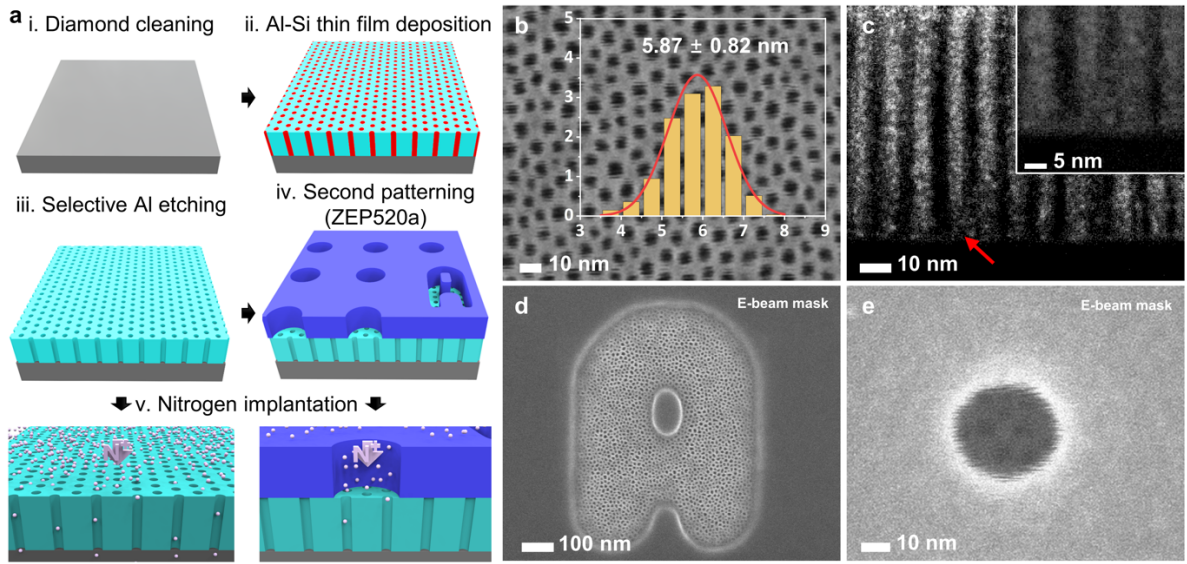

**Figure 1.** Fabrication process of nanoscale aperture arrays (NAAs), and a secondary electron beam lithography (EBL) mask. (a) Schematic of the overall process: i) cleaning of the diamond substrate by tri-acid boiling, ii) Al-Si thin film deposition by RF sputtering, iii) aluminum etching in acid solution, iv) secondary mask pattering by EBL, and v) nitrogen ion (14N+) implantation with and without a secondary mask. (b, c) Surface field emission scanning electron microscope (surface FE-SEM) and cross-sectional high-angle annular dark field– scanning transmission electron

microscopy (HAADF-STEM) images of NAAs on a diamond surface. (d, e) FE-SEM images of patterns formed by EBL on the NAAs. (e) Several NAAs could be observed through a single EBL hole.

Figure 1a shows a schematic of the NAA fabrication process. Single-crystal diamonds are tri-acid cleaned (step i). A phase-separated Al-Si thin film with a thickness of about 55 nm is then deposited by RF sputtering (step ii). Al nanowires are selectively etched, leaving only the NAAs layer comprising amorphous oxidized silicon (step iii) (Figure 1b). The optimized NAAs show a mean aperture size ($d_a$) of 5.87 nm and a mean separation wall width of 4.8 nm. The fully etched aperture is confirmed by the cross-sectional HAADF-STEM images shown in Figure 1c. Straight and vertical holes are clearly formed on the diamond. We then adjust the fabrication parameters and choose NAAs with $d_a$ values of 5.87 and 6.93 nm for samples B and C, respectively, to investigate the effects of the NAA size on the NV ensemble formation. To isolate several NAAs, we apply a secondary EBL mask with a circular hole pattern with a diameter of about 30 nm (Figure 1d-e) and a thickness of about 200 nm on one of the B samples, which is then labeled sample A (step iv). The mean EBL hole size on sample A is 32.23 nm. For use as a reference, we prepare a bare diamond without NAAs for ion implantation under the same conditions, and the diamond is labeled sample D.

Next, $^{14}N^+$ ions are implanted at 10 keV with an ion dose of $4 \times 10^{13}/cm^2$ by INNOVION Corp. for all four samples (step v). Implantation parameters are determined by the geometric conditions of the NAAs and EBL mask. Assuming that the [N] to [NV] conversion yield is below 3%, we aim to create a small number of NVs (less than four) per EBL hole spot with a reasonably long coherence time, which would satisfy strong coupling criteria. Here, the strong coupling regime is defined as $\frac{1}{v_{dip}} < T_{2,Hahn}$, where $v_{dip}$ is the average dipolar coupling between NV spins and $T_{2,Hahn}$ is the Hahn-echo spin coherence time of the NV. The 55 nm thickness of NAAs is sufficient to block ions accelerated at 10 keV, since the penetration depth calculated with Stopping and Range of Ions in Matter (SRIM)[28] is around 45 nm (projected range of 30.1 nm and longitudinal straggle of 15.3 nm). Finally, all the samples are annealed to generate NV centers, and an additional $O_2$ annealing process is performed to induce oxygen termination on the diamond surface for improved spin coherence time of shallow NV centers[29].

We measure the photoluminescence (PL) and Hahn-echo coherence time of ensemble NV centers in the aforementioned four fabricated diamond samples by using a homebuilt confocal microscope. Figure 2a shows the PL of ensemble NVs per laser spot area (spot diameter ~ 400 nm) for 0.1 mW power with a 532 nm excitation laser and their ensemble Hahn-echo spin coherence time ($T_{2,Hahn}$) for a bias magnetic field of 70 G. The red error bars are the standard deviation of $T_{2,Hahn}$ for each sample. The final sample, sample A, shows a larger deviation than the other incompletely fabricated samples, especially the bare diamond (sample D), owing to the imperfect mask fabrication technique employed.

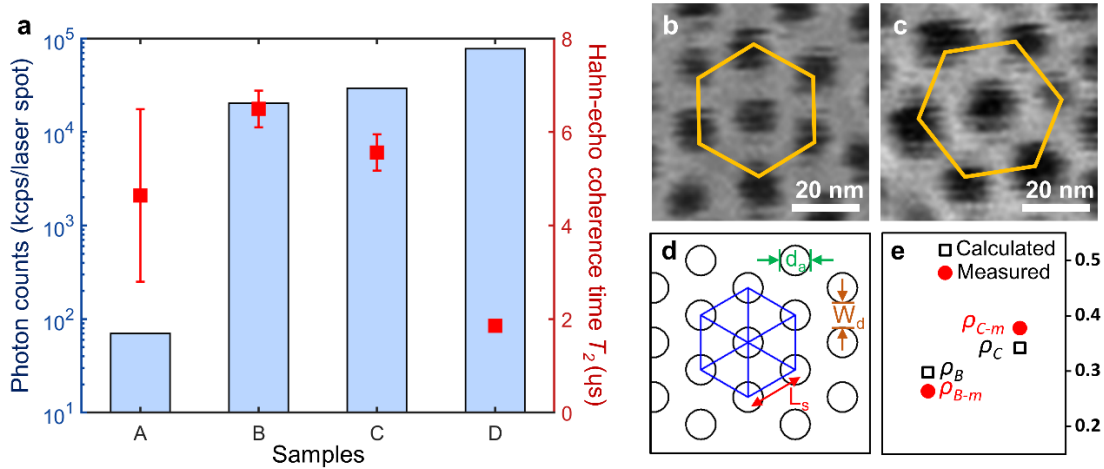

**Figure 2.** (a) Photoluminescence (PL) measured on a logarithmic scale (blue bars) and Hahn-echo spin coherence time (red squares) of ensemble NVs per laser spot size area for four diamond samples. (b-c) FE-SEM images of samples A and B. (d) A schematic of a perfectly hexagonal NAA. (e) Calculated and measured open area ratios of the NAAs.

The NVs in sample D show the highest PL and the shortest coherence time, while samples B and C show lower PL and longer coherence times. This observation indicates that the NAAs effectively reduce the NV density while increasing the average distance between NV and nitrogen electronic spin impurities[30]. Compared with sample B, the NVs in sample C show higher PL and a slightly shorter coherence time, implying denser nitrogen ion substitution owing to the larger aperture size (Fig. 2b and c). Since the measured PL is the sum of photon counts of NVs within the diffraction limit[31] and since the ensemble Hahn-echo coherence time is inversely proportional to the ion implantation dose[30], our measurement data show that the NAAs successfully act as a blocking mask despite the sub-10 nm scale aperture, and the number of implanted nitrogen ions is reduced by the open area ratio of the NAAs.

To quantitatively verify the NAAs' masking effect, we calculate the ratio between the aperture area with diameter $d_a$ and a unit hexagonal area and compare it with the ratio derived from PL measurement data. To simplify the calculation, we assume that NAAs comprised equal diameter apertures and a perfectly hexagonal array as shown in Fig. 2d. The ratio of the aperture area to a unit hexagonal area is given by

$$A_{Total} = \frac{3\sqrt{3}}{2}(L_s)^2 \qquad (1)$$

$$A_{apertures} = \pi\left(\frac{d_a}{2}\right)^2 \qquad (2)$$

$$\rho = (A_{apertures} * 3)/A_{Total} \qquad (3)$$

$L_s$ is the side length of the hexagon, and it is the sum of the hole diameter $d_a$ and the width of the wall, $W_d$, between apertures. The area ratios of samples B and C are calculated to be $\rho_B = 0.298$ and $\rho_C = 0.342$, respectively, based on FE-SEM measurement data. In Figure 2e, we compare the calculated area ratios with measured $\rho$ values, $\rho_{B-m} = 0.264$ and $\rho_{C-m} = 0.378$, derived from PL measurement ratio of sample B to that of sample D and the PL measurement ratio of sample C to that of sample D, assuming that [N] to [NV] conversion yields are identical for all samples. The slight discrepancy is due to the geometric differences in the NAAs' $L_s$ and $d_a$ and our crude assumption about the [N] to [NV] conversion yield which can differ for shallow implanted NVs.

Furthermore, the ensemble coherence time ($T_{2,Hahn}$) for the calculated effective implantation dose ($1.056 \times 10^{13}$ for sample B and $1.512 \times 10^{13}$ for sample C, obtained by multiplying $\varrho_{B-m}$ and $\varrho_{C-m}$ to the used dose of $4 \times 10^{13}$) agrees with that for the given implantation dose and energy reported in the literature[32], indicating that the NAAs successfully blocked the nitrogen ions.

Sample A, which is our final sample, shows lower PL per laser spot than sample B or C, with the PL being lower by a factor of about 280, which strongly indicates good localization of the cluster of NVs. This reduction is on the order of the magnitude range of the EBL mask's geometric confinement ratio, considering that the EBL hole to laser spot area ratio is about 200. Furthermore, the NV PL in sample A is slightly lower than expected, and we observe that the variation of $T_{2,Hahn}$ (Figure 2a) for sample A is greater than that for the other samples. We believe that these observations resulted from the E-beam resist remaining in the NAAs during the development process, which reduced the implantation energy of ions. Nevertheless, with the NAAs and EBL double-layered mask, we can clearly observe the cluster of NVs having expected spin coherence times.

We carefully perform optical verification of NV centers in sample A, which are ion implanted through 5.87 nm diameter NAAs confined within 30 nm secondary circular EBL hole patterns and conduct single-NV spin measurements. To optically resolve each cluster, we generate 2D grid hole pattern with 2 μm separation (Figure 3a). Confocal image reveal that about 70% of the holes are empty, yet more than one NVs are observed in the remaining holes. To confirm the statistical distribution of NVs per spot, we measure PL, the second-order autocorrelation function ($g^{(2)}$), and the ODMR. First, we approximately determine the number of NVs per spot by dividing the total PL count by the average single-NV photon count (~60 kcps). Next, by measuring the $g^{(2)}(t)$, we refine statistical counts from the previous PL estimates (Figure 3d–f). $g^{(2)}(t)$ shows an antibunching dip at t = 0, and the number of single photon sources N is deduced from the equation $g^{(2)}(0) = (1-1/N)$ [33], where $g^{(2)}(0)$ dip values are extracted from the $g^{(2)}(t)$ fit to data; $g^{(2)}(0) < 0.5$ means a single NV, $0.5 < g^{(2)}(0) < 0.66$ indicates two NVs, and $0.66 < g^{(2)}(0) < 0.75$ implies three NVs.

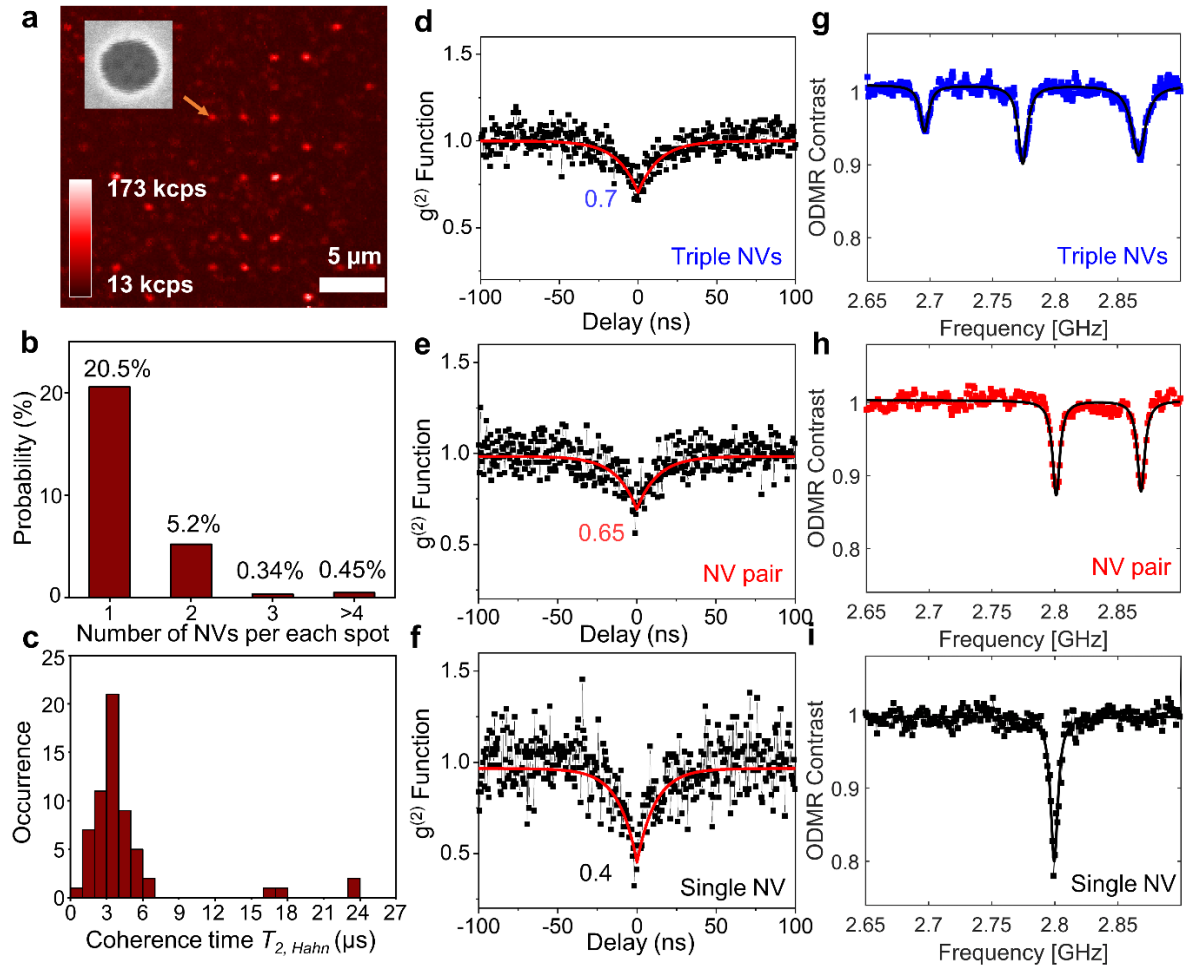

**Figure 3.** (a) Confocal PL image of NVs in sample A. Typically, 5–6 NAAs are observed per EBL hole. (b) Overall extracted probability (for all EBL holes) of the number of NVs formed per each hole spot in sample A. (c) Distribution of Hahn-echo coherence time ($T_{2,Hahn}$) measurements of a single NV center at different spots. (d–f) Autocorrelation function ($g^{(2)}(t)$) measurements at different EBL hole spots. The red lines are the best fits. By extracting $g^{(2)}(0)$ from the fit, we can estimate the number of NVs. (g–i) Optically detected magnetic resonance (ODMR) signals at different EBL holes. The external magnetic field is misaligned with all four crystalline axis and the number of ODMR dips

indicate the number of NVs per spot. Based on the ODMR results and previous optical measurements, we distinguish up to 3 NVs per EBL hole.
Therefore, in our measurements, $g^{(2)}(0) = 0.4$ denotes a single NV, 0.65 hints at two NVs, and 0.7 points to three NVs. We also measure ODMR (Figure 3g–i), with the external bias magnetic field intentionally misaligned with all four crystalline axis, and cross-correlate the number of ODMR dips with previous measurements; finally, we determine our statistical counts (Figure 3b). We clearly confirm up to three NVs per spot (Figure 3g).

In addition, we measure the Hahn-echo coherence time of a single NV at each hole, this time with a bias magnetic field of 70 G carefully aligned with the NV axis (Figure 3c). With some variation due to different local bath environments[31], 10% of the measured $T_{2,Hahn}$ exceed 16 µs, which are in a range comparable to the target coherence time of 20 µs. This range is important since the minimum coherence time required to observe relatively strong NV-NV dipolar coupling, assuming median angular contribution between two different crystalline axis NVs with a separation distance of 10 nm, is around 20 µs. The low average $T_{2,Hahn}$ of 4.5 µs is presumably because of the shallow NV implantation effect[34], which is led by ions being scattered from the remaining e-beam resist layer. This problem can be resolved by using an SOI wafer mask and a transfer method[13], implantation with the pierced AFM tip method[19], or combining EBL with a dry-etching process[24].

**Table 1.** Comparison of device engineering capabilities of different fabrication schemes.

|  | **Molecular**[21] | **AFM tip**[12] | **EBL Au**[24] | **EBL+ALD**[13] | **This work** |
|---|---|---|---|---|---|
| **Precision (area)** | N/A | ~700 nm² | 78 nm² | 100 nm² | 28 nm² |
| **Coupling** | Good (<10 nm) | Poor (~30 nm) | Poor (~40 nm) | Moderate (~16 nm) | Moderate (~10 nm) |
| $T_{2,Hahn}$ | ~300 µs | N/A | ~10 µs | N/A | ~4 µs |
| **Uniformity** | Poor | Good | Moderate | Moderate | Good |
| **Scalability** | Poor | Moderate | Good | Moderate | Good |

In terms of the precise spin defect generation in a solid-state system, the most important figure of merit is the opening area of ion implantation masks. As summarized in Table 1, our work records by far the smallest single aperture mask opening area of 28 nm² and the closest center-to-center width of about 10 nm ever obtained, which could cause nearest neighboring spins to be strongly coupled to each other. In contrast to other schemes that require careful fabrication of aperture arrangements, NAAs are by their nature orderly arranged with triangular lattice geometry and have an almost perfect success rate because of the homogeneous phase-separation process. This offers a huge advantage for generating uniformly and closely separated spin cluster systems, which we discuss next by referred to simulations.

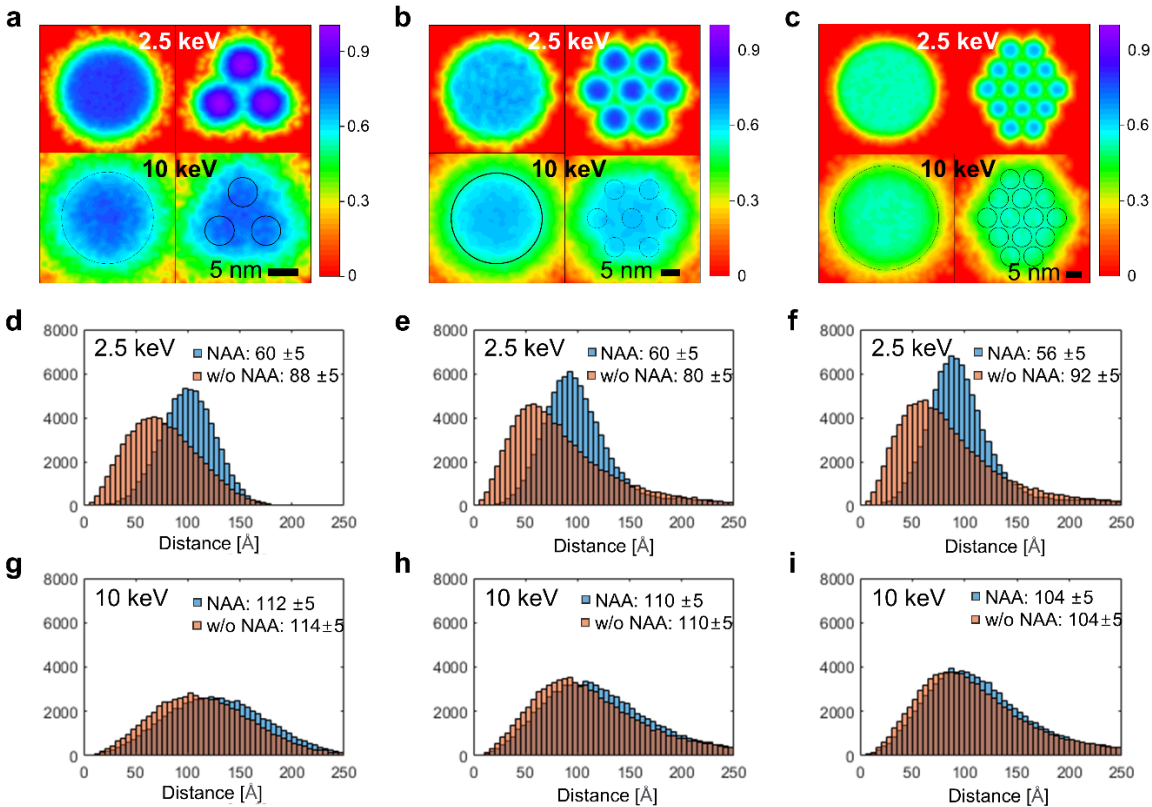

**Figure 4.** (a–c) Two-dimensional positional distribution of implanted ions relative to a single EBL hole spot (left) and an EBL hole combined with 6 nm NAAs (right) with implantation energies of 2.5 keV (top) and 10 keV (bottom),

calculated by SRIM and plotted as kernel density estimates. The palettes shown are kernel density values (normalized density probability distribution). The EBL aperture sizes are 18, 27, and 41 nm for a, b, and c, respectively. (d–i) Corresponding 3D nearest defect distance distributions with their full width at half-maximum (FWHM) values for different energies and sizes of the EBL hole[35].

We perform simulations, in which the average distance between generated NV spins is calculated, by varying the implantation energy and number of NAAs per hole to discuss scalability of the NAA-EBL double-layer mask technique. Figure 4 shows a 2D kernel density plot of implanted ions relative to a single EBL hole and several NAAs in the EBL hole simulated by SRIM[28].

From the SRIM simulation results, in the low implantation energy range under 3 keV, we expect to achieve deterministic control of the NV position and uniformity in the separation distance. When the implantation energy is set to 10 keV, distinct separation between ions through NAAs is barely observed owing to the relatively high lateral straggle (38 Å) compared with the distance between apertures (43 Å). For a lower implantation energy of 2.5 keV, the density of confined ions within the aperture area increased, showing relatively clear separation between NAAs (Figure 4a–c). Furthermore, as the number of NAAs within a hole spot increases, the variation in the nearest NV-NV separation distance remains well below the apertures' center-to-center distance of about 10 nm. However, without NAAs, the distance distribution gradually gets broaden, almost reaching the center-to-center distance (Figure 4d–f).

Although the NAA-EBL double-layer method shows promising results in terms of scalability, the low-energy implantation in general degrades the conversion yield and spin properties of NV centers. Recent works, however, have shown considerable progress in the enhancement of the conversion yield and spin properties; for example, the conversion yield of implanted nitrogen ions can be improved by laser writing[36], using a co-implantation technique[37], repeated annealing[38], and other types of impurities doping before NV formation[39]. To improve the spin properties of shallow

NVs, we can re-grow a diamond layer through microwave-assisted chemical vapor deposition to eliminate the diamond surface defect noise and increase the coherence time[34].

To summarize, by depositing eutectic Al-Si thin film directly on a diamond surface and etching only Al away, we fabricated 6 nm sized NAAs with a high aspect ratio and verified that NAAs could be used as an ion implantation blocking mask. By using the secondary EBL hole mask, we successfully confine clusters of NVs in EBL holes during the implantation process and confirmed up to three spins within a 30 nm diameter hole spot by cross-correlating optical and spin resonance measurements. Our NAA-EBL double-layer mask method is a promising tool for engineering qubits through deterministic positioning in a solid-state-defect-based qubit system, and it is especially robust for the fabrication of a large number of quantum nodes in a wide area. We believe that the geometric shape of the qubit cluster can be controlled by carefully designing the secondary mask pattern on NAAs. A combination of the NAA-EBL double-layer technique with better conversion yield and improved spin coherence time would enable strongly and uniformly interacting spin qubits for generating non-classical collective states. In particular, strongly interacting quasi-2D spin clusters can be engineered, and this system can be exploited as a universal tool for quantum sensing or quantum simulation applications.

**Methods**

**Sample fabrication.** Diamond substrates had [B] < 1 ppb, [N] < 5 ppb with less than 0.03 ppb intrinsic NV concentration. They were cleaned by >1 h at 170 °C in a mixture of sulfuric acid, perchloric acid, and nitric acid. On a cleaned diamond subtrate, 55 nm thick Al-Si phase separated thin film was deposited by RF sputtering. To minimize the nanoscale aperture size, we optimized phase-separated Al-Si thin film formation by controlling RF sputter conditions and found optimal condition of Ar working pressure 0.3 mTorr, substrate temperature 100 °C with RF power 150W at deposition rate of 15.8 nm/min for our final device. The Al nanowires were selectively etched by immersing the samples in 5% phosphoric acid for overnight leaving only NAAs layer, composed with amorphous silicon oxide. After the implantation, all the samples were annealed at 800 °C for 8 h and at 1100 °C for 2 h to generate NV centers. Additional $O_2$ annealing process was performed at 450 °C for 4 h.

**Optical and spin measurements**. Optical and spin measurements were done using a home-built confocal scanning laser microscope. An acousto-optic modulator (Gooch&Housego 3200-121) allowed time-gating of a 300 mW 532 nm diode-pumped solid state laser (MGL-III-532-300mW). Here, double-path method was used to enhance isolation ratio, and the 1st order diffraction beam was coupled to a single-mode fiber. Reflected after a long-pass filter, fiber coupled green beam passed through an air objective (Olympus UPLXAPO60XO), which then focused the beam onto an NV center. Diamond sample was fixed on a three-axis piezo controlled stage (P-562.3CD), and a copper wire was placed on top of a diamond for microwave signal delivery. NV red fluorescence signal was collected back through the same objective, then onto two silicon avalanche photodetectors (Perkin Elmer SPCM-ARQH-12) split by a 50:50 beam splitter. Finally, we place a pinhole (diameter 50$\mu$m) with f = 100 mm telescope and removed background due to unfocused light. All the measurements were carefully time-synchronized using pulse generating FPGA system with clock rate of 100MHz. All the measurements were performed at room temperature.

**Data availability**

All data supporting the findings of this study are available from the corresponding author upon reasonable request.

**Acknowledgment**

This work was supported by the National Research Foundation of Korea (NRF) (2019M3E4A1079777, 2019M3E4A1078660), the Institute for Information and Communications Technology Promotion (IITP) (2020-0-00947, 2020-0-00972), and the KIST research program (2E31021).

**Author Contributions**

T.-Y. Hwang, J. Lee, Y.-H. Choa and H. Jung conceived the research idea and T.-Y. Hwang and J. Lee designed the experiment. T.-Y. Hwang and S.-W. Jeon prepared the sample and performed the structural characterization. T.-Y. Hwang and J. Lee performed the optical and spin characterizations. T.-Y. Hwang, J. Lee and H. Jung wrote the manuscript. All authors contributed to the discussion and manuscript revision.

**Competing interests**

The authors declare no competing interests.

**References**

1. S. Hong, M. S. Grinolds, L. M. Pham, D. Le Sage, L. Luan, R. L. Walsworth, and A. Yacoby, Nanoscale magnetometry with NV centers in diamond, MRS Bulletin **38**, 155–161 (2013).

2. K. Arai, C. Belthangady, H. Zhang, N. Bar-Gill, S. J. DeVience, P. Cappellaro, A. Yacoby, and R. L. Walsworth, Fourier magnetic imaging with nanoscale resolution and compressed sensing speed-up using electronic spins in diamond, *Nature Nanotechnology* **10**, 859–864 (2015).

3. B. Hensen, H. Bernien, A. E. Dréau, A. Reiserer, N. Kalb, M. S. Blok, J. Ruitenberg, R. F. L. Vermeulen, R. N. Schouten, C. Abellán, W. Amaya, V. Pruneri, M. W. Mitchell, M. Markham, D. J. Twitchen, D. Elkouss, S. Wehner, T. H. Taminiau, and R. Hanson, Loophole-free Bell inequality violation using electron spins separated by 1.3 kilometres, *Nature* **526**, 682–686 (2015).

4. P. Neumann, R. Kolesov, B. Naydenov, J. Beck, F. Rempp, M. Steiner, V. Jacques, G. Balasubramanian, M. L. Markham, D. J. Twitchen, S. Pezzagna, J. Meijer, J. Twamley, F. Jelezko, and J. Wrachtrup , Quantum register based on coupled electron spins in a room-temperature solid, *Nature Physics* **6**, 249–253 (2010).

5. K. Nemoto, M. Trupke, S. J. Devitt, B. Scharfenberger, K. Buczak, J. Schmiedmayer, and W. J. Munro, Photonic quantum networks formed from NV⁻ centers, *Scientific Reports* **6**, 26284 (2016).

6. P. Neumann, J. Beck, M. Steiner, F. Rempp, H. Fedder, P. R. Hemmer, J. Wrachtrup, and F. Jelezko, Single-shot readout of a single nuclear spin, *Science* **329**, 542–544 (2010).

7. T. H. Taminiau, J. Cramer, T. van der Sar, V. V. Dobrovitski, and R. Hanson, Universal control and error correction in multi-qubit spin registers in diamond, *Nature Nanotechnology* **9**, 171–176 (2014).

8. S. Zaiser, T. Rendler, I. Jakobi, T. Wolf, S.-Y. Lee, S. Wagner, V. Bergholm, T. Schulte-Herbrüggen, P. Neumann, and J. Wrachtrup, Enhancing quantum sensing sensitivity by a quantum memory, *Nature Communications* **7**, 12279 (2016).

9. T. Pompili, S. L. N. Hermans, S. Baier, H. K. C. Beukers, P. C. Humphreys, R. N. Schouten, R. F. L. Vermeulen, M. J. Tiggelman, L. Dos Santos Martins, B. Dirkse, S. Wehner, and R. Hanson, Realization of a multinode quantum network of remote solid-state qubits, *Science* **372**, 259 (2021).

10. N. Y. Yao, L. Jiang, A. V. Gorshkov, P. C. Maurer, G. Giedke, J. I. Cirac, and M. D. Lukin, Scalable architecture for a room temperature solid-state quantum information processor, *Nature Communications* **3**, 800 (2012).

11. M. J. Degen, S. J. H. Loenen, H. P. Bartling, C. E. Bradley, A. L. Meinsma, M. Markham, D. J. Twitchen, and T. H. Taminiau, Entanglement of dark electron-nuclear spin defects in diamond, *Nature Communications* **12**, 3470 (2021).

12. J. Riedrich-Möller, S. Pezzagna, J. Meijer, C. Pauly, F. Mücklich, M. Markham, A. M. Edmonds and C. Becher, Nanoimplantation and Purcell enhancement of single nitrogen-vacancy centers in photonic crystal cavities in diamond, *Applied Physics Letters* **106**, 221103 (2015).

13. I. Bayn, E. H. Chen and M. E. Trusheim, L. Li, T. Schröder, O. Gaathon, M. Lu, A. Stein, M. Liu, K. Kisslinger, H. Clevenson, and D. Englund, Generation of ensembles of individually resolvable nitrogen vacancies using nanometer-scale apertures in ultrahigh-aspect ratio planar implantation masks, *Nano Letters* **15**, 1751–1758 (2015).

14. C. E. Bradley, J. Randall, M. H. Abobeih, R. C. Berrevoets, M. J. Degen, M. A. Bakker, M. Markham, D. J. Twitchen, and T. H. Taminiau, A ten-qubit solid-state spin register with quantum memory up to one minute, *Physical Review X* **9**, 031045 (2019).

15. G. Kucsko, S. Choi, J. Choi, P. C. Maurer, H. Zhou, R. Landig, H. Sumiya, S. Onoda, J. Isoya, F. Jelezko, E. Demler, N. Y. Yao, and M. D. Lukin, Critical thermalization of a disordered dipolar spin system in diamond, *Physical Review Letters* **121**, 023601 (2018).

16. S. Choi, J. Choi, R. Landig, G. Kucsko, H. Zhou, J. Isoya, F. Jelezko, S. Onoda, H. Sumiya, V. Khemani, C. von Keyserlingk, N. Y. Yao, E. Demler, and M. D. Lukin, Observation of discrete time-crystalline order in a disordered dipolar many-body system, *Nature* **543**, 221–225 (2017).

17. F. Dolde, I. Jakobi, B. Naydenov, N. Zhao, S. Pezzagna, C. Trautmann, J. Meijer, P. Neumann, F. Jelezko, and J. Wrachtrup, Room-temperature entanglement between single defect spins in diamond, *Nature Physics* **9**, 139–143 (2013).

18. J. Meijer, B. Burchard, M. Domhan, C. Wittmann, T. Gaebel, I. Popa, F. Jelezkoa, and J. Wrachtrup, Generation of single colour centres by focused nitrogen implantation, *Applied Physics Letters* **87**, 261909 (2005).

19. N. Raatz, C. Scheuner, S. Pezzagna and J. Meijer, Investigation of ion channeling and scattering for single-ion implantation with high spatial resolution, *Physica Status Solidi A. Applications and Materials Science* **216**, 1–13 (2019).

20. S. Sangtawesin, T. O. Brundage, Z. J. Atkins, and J. R. Petta, Highly tunable formation of nitrogen-vacancy centers via ion implantation, *Applied Physics Letters* **105**, 063107 (2014).

21. M. Haruyama, S. Onoda, T. Higuchi, W. Kada, A. Chiba, Y. Hirano, T. Teraji, R. Igarashi, S. Kawai, H. Kawarada, Y. Ishii, R. Fukuda, T. Tanii, J. Isoya, T. Ohshima, and O. Hanaizumi, Triple nitrogen-vacancy centre fabrication by $C_5N_4H_n$ ion implantation, *Nature Communications* **10**, 4–9 (2019).

22. T. Gaebel, M. Domhan, I. Popa, C. Wittmann, P. Neumann, F. Jelezko, J. R. Rabeau, N. Stavrias, A. D. Greentree, S. Prawer, J. Meijer, J. Twamley, P. R. Hemmer, and J. Wrachtrup, Room-temperature coherent coupling of single spins in diamond, *Nature Physics* **2**, 408–413 (2006).

23. I. Jakobi, S. A. Momenzadeh, F. F. De Oliveira, J. Michl, F. Ziem, M. Schreck, P. Neumann, A. Denisenko, and J. Wrachtrup, Efficient creation of dipolar coupled nitrogen-vacancy spin qubits in diamond, *Journal of Physics: Conference Series* **752**, 012001 (2016).

24. D. Scarabelli, M. Trusheim, O. Gaathon, D. Englund, and S. J. Wind, Nanoscale engineering of closely-spaced electronic spins in diamond, *Nano Letters* **16**, 4982–4990 (2016).

25. R. Fukuda, P. Balasubramanian, I. Higashimata, G. Koike, T. Okada, R. Kagami, T. Teraji, S. Onoda, M. Haruyama, and K. Yamada, Lithographically engineered shallow nitrogen vacancy centers in diamond for external nuclear spin sensing, *New Journal of Physics* **20**, 083029 (2018).

26. K. Fukutani, K. Tanji, T. Motoi and T. Den, Ultrahigh pore density nanoporous films produced by the phase separation of eutectic Al–Si for template-assisted growth of nanowire array, *Advanced Materials* **16**, 1456–1460 (2004).

27. K. Fukutania, K. Tanji, T. Saito and T. Den, Phase-separated Al–Si thin films, *Journal of Applied Physics* **98**, 033507 (2005).

28. J. P. Biersack and L. G. Haggmark, A Monte Carlo computer program for the transport of energetic ions in amorphous targets, *Nuclear Instruments and Methods* **174**, 257–269 (1980).

29. S. Sangtawesin, B. L. Dwyer, S. Srinivasan, J. J. Allred, L. V. H. Rodgers, K. De Greve, A. Stacey, N. Dontschuk, K. M. O'Donnell, D. Hu, D. A. Evans, C. Jaye, D. A. Fischer, M. L. Markham, D. J. Twitchen, H. Park, M. D. Lukin, and N. P. de Leon, Origins of diamond surface noise probed by correlating single-spin measurements with surface spectroscopy, *Physical Review X* **9**, 1–17 (2019).

30. E. Bauch, S. Singh, J. Lee, C. A. Hart, J. M. Schloss, M. J. Turner, J. F. Barry, L. M. Pham, N. Bar-Gill, S. F. Yelin, and R. L. Walsworth, Decoherence of ensembles of nitrogen-vacancy centers in diamond, *Physical Review B* **102**, 134210 (2020).

31. S. Pezzagna, B. Naydenov, F. Jelezko, J. Wrachtrup, and J. Meijer, Creation efficiency of nitrogen-vacancy centres in diamond, *New Journal of Physics* **12**, 065017 (2010).

32. J.-P. Tetienne, R. W. de Gille, D. A. Broadway, T. Teraji, S. E. Lillie, J. M. McCoey, N. Dontschuk, L. T. Hall, A. Stacey, D. A. Simpson, and L. C. L. Hollenberg, Spin properties of dense near-surface ensembles of nitrogen-vacancy centers in diamond, *Physical Review B* **97**, 085402 (2018).

33. M. B. Plenio and P. L. Knight, The quantum-jump approach to dissipative dynamics in quantum optics, *Review of Modern Physics* **70**, 101 (1998).

34. T. Staudacher, F. Ziem and L. Häussler, R. Stöhr, S. Steinert, F. Reinhard, J. Scharpf, A. Denisenko, and J. Wrachtrup, Enhancing the spin properties of shallow implanted nitrogen vacancy centers in diamond by epitaxial overgrowth, *Applied Physics Letters* **101**, 212401 (2012).

35. O. Lehtinen, B. Naydenov, P. Börner, K. Melentjevic, C. Müller, L. P. McGuinness, S. Pezzagna, J. Meijer, U. Kaiser, and F. Jelezko, Molecular dynamics simulations of shallow nitrogen and silicon implantation into diamond, *Physical Review B* **93**, 035202 (2016).

36. Y-C. Chen, B. Griffiths, L. Weng, S. S. Nicley, S. N. Ishmael, Y. Lekhai, S. Johnson, C. J. Stephen, B. L. Green, G. W. Morley, M. E. Newton, M. J. Booth, P. S. Salter, and J. M. Smith, Laser writing of individual nitrogen-vacancy defects in diamond with near-unity yield, *Optica* **6**, 662 (2019).

37. B. Naydenov, V. Richter, J. Beck, M. Steiner, P. Neumann, G. Balasubramanian, J. Achard, F. Jelezko, J. Wrachtrup, and R. Kalish, Enhanced generation of single optically active spins in diamond by ion implantation, *Applied Physics Letters* **96**, 163108 (2010).

38. S. Chakravarthi, C. Moore, A. Opsvig, C. Pederson, E. Hunt, A. Ivanov, I. Christen, S. Dunham, and K-M. C. Fu, Window into NV center kinetics via repeated annealing and spatial tracking of thousands of individual NV centers, *Physical Review Materials* **4**, 023402 (2020).

39. T. Lühmann, R. John, R. Wunderlich, J. Meijer, and S. Pezzagna, Coulomb-driven single defect engineering for scalable qubits and spin sensors in diamond, *Nature Communications* **10**, 4956 (2019).